**This is the accepted manuscript (postprint) of the following article:**
E. Salahinejad, M. Jafari Baghjeghaz, *Structure, biomineralization and biodegradation of Ca-Mg oxyfluorosilicates synthesized by inorganic salt coprecipitation*, Ceramics International, 43 (2017) 10299-10306.
https://doi.org/10.1016/j.ceramint.2017.05.059

# Structure, biomineralization and biodegradation of Ca-Mg oxyfluorosilicates synthesized by inorganic salt coprecipitation

E. Salahinejad *, M. Jafari Baghjeghaz

Faculty of Materials Science and Engineering, K.N. Toosi University of Technology, Tehran, Iran

## Abstract

In this research, a novel group of Ca-Mg oxyfluorosilicates containing different levels of fluoride substituting for oxide was synthesized by an inorganic salt coprecipitation process followed by calcination/sintering. The effects of the incorporation of fluoride on the resultant structural characteristics, apatite-forming ability and biodegradability were evaluated by X-ray diffraction, transmission electron microscopy, scanning electron microscopy/energy-dispersive X-ray spectroscopy, Fourier transform infrared spectroscopy, inductively coupled plasma spectroscopy and pH measurements. According to the results, the samples containing up to 2 mol% F present a single-phase structure of diopside ($MgCaSi_2O_6$) doped with F. It was also found that to meet the most biomineralization characteristic, the optimal value of fluoride in the homogeneous samples is 1 mol%. In this regard, on the one hand, the partial incorporation of fluoride into apatite (via forming fluorohydroxyapatite) and, on the other hand, the absence of fluorite ($CaF_2$) as a consumer of Ca in the deposits are responsible for achieving the most apatite-forming ability circumstance controlled by an ion-exchange reaction mechanism. In conclusion, this study reflects the merit of the optimization of fluoride-doping into Ca-Mg silicates for development in biomedicine.

* Corresponding Author:
Email Addresses: <salahinejad@kntu.ac.ir>

## 1. Introduction

Ceramic materials are currently considered in biomedicine, particularly in dentistry. For instance, as the classic members of bioinert ceramics, alumina and zirconia are principal constitutes in fabricating dental prostheses. Bioactive glasses and glass-ceramics are also used in air polishing, toothpastes, bonding agents and tooth regeneration purposes [1, 2]. Typically, their bioactive feature in the toothpastes limits dentine hypersensitivity via the precipitation of apatite on the tooth's surface and sealing dentinal tubules [3, 4]. Additionally, porous ceramics like calcium phosphate-based materials have been used to fill tooth and bone defects. Some bioceramics like calcium silicates have been also tested in dentistry and endodontic therapy as root repair materials and for apical retrofills. In the recent examples, the advantage of using the ceramics from the viewpoint of bioactivity is their ability to form apatite and to create a chemical bond to the dentin [5].

Diopside with the chemical formula $MgCaSi_2O_6$ is a bioactive ceramic of the $SiO_2$-CaO-MgO system. Because of its suitable thermal expansion coefficient and bond strength to biometals, this silicate has been used as a coating material for orthopedic and dental applications [6-8]. Also, glass-ceramics with the diopside composition have been successfully tested as a veneering species for restoration purposes. Despite the relatively acceptable bioactivity of diopside which justifies its use in dental root implants [9], it seems that its more development demands more improvements in bioactivity.

Traditionally, the incorporation of fluoride is considered in dental resins and glass ionomer cements to enhance the remineralization of filling materials [10]. This ion can remineralize enamel and softened dentine, thereby decreasing the formation of caries [11].

Also, fluoride induces the precipitation of fluorapatite on teeth, challenging their demineralization and degradation [12-14]. In this research, a group of F-doped diopside-based bioceramics (Ca-Mg oxyfluorosilicate) was synthesized by a coprecipitation process using inorganic salt precursors, which has been successfully tested to prepare other multi-oxide ceramics [15-23]. Then, to meet a better bioactivity, the optimization of fluoride-doping into the samples was regarded.

## 2. Materials and methods

For the coprecipitation synthesis of the stoichiometric Ca-Mg silicate (diopside, $MgCaSi_2O_6$) named 0F, an equimolar content of calcium chloride ($CaCl_2$, Merck, >98%) and magnesium chloride ($MgCl_2$, Merck, >98%) and a double-molar amount of silicon tetrachloride ($SiCl_4$, Merck, >99%) were dissolved in dry ethanol ($C_2H_5OH$, Merck, >99%). Aqueous ammonia solution ($NH_4OH$, Merck, 25%) was then added to the above solution to obtain pH of 10, giving white precipitates. For the addition of fluoride, 5, 10, 25, 50 and 100 mol% of $MgCl_2$ were replaced with the same molar amounts of magnesium fluoride ($MgF_2$, Alfa Aesar, >99%), keeping the molar ratio of Ca/Mg/Si = 1/1/2. These samples were denominated 1F, 2F, 3F, 4F and 5F, respectively, yielding about 1, 2, 5, 10, 20 mol% F in the final products.

The synthesized precipitates were dried, calcined, and then analyzed by X-ray diffraction (XRD). From the XRD data, after the identification of the formed phases, the crystallite sizes and lattice variables (lattice parameters, interaxial angle and lattice volume) of the diopside phase were determined by the Scherrer formula and the crystallographic relations of the monoclinic system, respectively. Also, one of the prepared powder samples was selectively analyzed by transmission electron microscopy (TEM) to characterize the particle size.

The homogenous diopside samples, according to the XRD analyses, were sintered at 1200 °C for 2 h and then soaked in the simulated body fluid (SBF) [24] at 37 °C for different durations. The immersed samples were then studied by Field-emission scanning electron microscopy (FESEM)/energy-dispersive X-ray spectroscopy (EDS) and Fourier transform infrared spectroscopy (FTIR). Also, the SBF after immersion of the samples was analyzed for Si, Mg, Ca and P by inductively coupled plasma spectroscopy (ICP). The pH value of the SBF after the incubation of the samples was also measured to further assess the materials' *in vitro* biodegradation and bioactivity.

## 3. Results and discussion

Fig. 1 depicts the XRD pattern of the 0F, 1F, 2F, 3F, 4F and 5F samples after calcination. Based on the analysis of the profiles by the PANalytical X'Pert HighScore program, the 0F, 1F and 2F samples present a single-phase structure of diopside ($MgCaSi_2O_6$, 00-017-0318) with a monoclinic crystalline structure. Nonetheless, in the 3F, 4F and 5F samples, fluorite ($CaF_2$, 00-004-0864), cuspidine ($Ca_4Si_2O_7(F,OH)_2$, 01-085-1334), akermanite ($Ca_2MgSi_2O_7$, 00-035-0592) and monticellite ($CaMgSiO_4$, 00-035-0590) phases are detected as well as diopside, whereas the 5F sample is closely diopside-free. Indeed, by increasing the incorporation of fluoride beyond 10 % of the precursor substitution, the amount of diopside in the calcined structure is reduced; alternatively, the other phases, particularly the F-containing ones (fluorite and cuspidine), appear. That is, the saturation/supersaturation solubility limit of fluoride in diopside is almost 2 mol% which is related to the 2F sample, based on the XRD analysis. Considering the electric charges of the species, the incorporation of fluoride (two ions) in diopside is conducted via substitution for oxygen (one ion), replacing bridging oxide bonds with nonbridging fluoridated ones [25, 26], as schematically drawn in Fig. 2.

The lattice parameters, interaxial angle, lattice volume and crystallite size of diopside, extracted from the XRD data, are listed in Table 1. The 5F sample is not involved in this table because it is nearly diopside-free, as demonstrated by the XRD analysis. Diopside is crystallized at a monoclinic system with the lattice parameters of $a \neq b \neq c$ and the interaxial angles of $\alpha = \gamma = 90 \neq \beta$, where its lattice volume ($V$) equals to $abc \sin \beta$. The lower lattice volume of the undoped sample (0F) comprising nanoparticles of almost 70 nm in diameter (Fig. 3), in comparison to the theoretical value [27], is attributed to the presence of crystal imperfections in the lattice. It is seen that by increasing the fluoride incorporation from 0F to 2F, where homogenous diopside was detected by XRD, the lattice volume is enhanced. This is due to the fact that each oxygen anion is replaced with two fluoride anions (Fig. 2), expanding the lattice. Also, in the same range of the fluoride addition, the crystallite size of diopside is decreased. Because at a given firing temperature, a higher homologous temperature and thereby higher crystallite growth are experienced by increasing the fluoride content, since fluoride reduces the melting point of silicates [28, 29]. However, by increasing the fluoride inclusion from 2F to 4F, in which the samples are multiphase, on the one side, the lattice volume of fluoridated diopside is reduced; and on the other side, its crystallite size shows an increasing trend. Both of them can be owing to the depletion of diopside from fluoride, so that this anion tends to leave diopside toward the other phases existing in the samples, probably fluorite and cuspidine due to differences in chemical potential, thereby annihilating the saturation/supersaturation of diopside from fluoride. The nearness of the lattice variables of the 5F to those of the F-free sample (0F) can be another evidence for this conclusion. In this regard, the accumulation of fluoride in cuspidine and fluorite has been previously reported in the thermal crystallization of some fluoride-containing bioactive glasses [25, 30].

Only the homogeneous samples, i.e. 0F, 1F and 2F, were studied in terms of *in vitro* bioactivity and biodegradation. The SEM micrographs and EDS patterns taken of the surface of the sintered samples after 3, 7 and 14 days of immersion in the SBF are indicated in Figs. 4, 5 and 6, respectively. In accordance to Fig. 4a, the majority of the 0F surface after soaking in the SBF for 3 days reflects no typical apatite-like precipitates; only nanometric-sized precipitates are observable on a small part of the surface in the higher-magnification micrograph (Fig. 4b). Comparatively, the amount of the apatite-like precipitates on the 1F surface has been enhanced, with a noncontinuous rose-like feature (Figs. 4d and 4e). In contrast, this type of precipitates are hardly recognized on the 2F surface in Figs. 4g and 4h. Concerning 7 days of incubation in the SBF, some rose-like precipitates appear on 0F (Figs. 5a and 5b), whereas the amount of the precipitates for 1F has been increased than 3 days and is still maximum among all of the samples (Figs. 5d and 5e). Also, at the same incubation period, the 2F surface presents a number of new discrete nano-sized plates, as provided in Figs. 5g and 5h. A similar ranking for biomineralization in the SBF was also found for 14 days of incubation (Fig. 6), so that the 1F surface is continuously covered by the rose-like deposits. From the comparison of the above-described micrographs, it was summarized that the *in vitro* mineralization increases in the ranking of 2F, 0F and 1F for each soaking period and with immersion time from 3 days to 14 days for each sample.

The EDS patterns taken of the precipitates deposited as a result of soaking in the SBF (the c, f and i parts of Figs. 4, 5 and 6) demonstrate the presence of phosphorous (P) on the samples' surfaces. It suggests that the formed precipitates are P-containing and a type of apatite, since the samples were essentially P-free. Note that to record high-intensity EDS signals, a high-voltage of 15 kV had been used. This led to the appearance of signals of the samples, as well as the precipitates, due to the small dimensions of the latter. The samples (0F, 1F and 2F) are composed of Si, Mg, Ca, O and F, while the apatite-like deposits contain Ca and P. It means that the appearance of Si and Mg in the EDS spectra is merely due to the signals received from the samples, whereas the P signals are only related to the precipitates. Thus, the Ca peaks originate from both the samples and precipitates. Accordingly, the intensity ratio of P/Si is employed as a criterion to indirectly compare the level of the apatite-like deposits. The higher amount of this parameter dictates that higher-intensity signals of the deposits and/or lower-intensity signals of the samples are received, suggesting the formation of a higher level of apatite on the surface. The P/Si parameters for the samples soaked in the SBF for the different periods, extracted from the EDS patterns, are signified in Fig. 7. It infers that the most- and less-effective biomineralization abilities comparatively belong to 1F and 2F, respectively. Also, the apatite formation is increased with the immersion time, confirming the SEM observations of Figs. 4, 5 and 6.

The FTIR spectra of the samples after immersion in the SBF for 14 days are represented in Fig. 8. The peaks appearing at about 470, 525, 965 and 1075 $cm^{-1}$ are signified to phosphate in all of the incubated samples, suggesting that the precipitates formed in the SBF are a kind of apatite (calcium phosphate). An additional absorption mode of phosphate is also detected at about 1020 $cm^{-1}$ for 0F and 1F, whereas this vibration does not fairly exist for 2F. This, like SEM/EDS, verifies the lower biomineralization of 2F in comparison to 0F and 1F. Also, the higher intensity of the vibrations corresponding to the phosphate group for 1F is indicative of its higher apatite-forming ability, compared with 0F. That is, the FTIR analysis confirms the biomineralization ranking of 1F > 0F > 2F, realized by the SEM and EDS studies. A small signal of the hydroxyl group is recognized at around 1645 $cm^{-1}$, suggesting that the formed deposits are essentially the hydroxyapatite. The lower intensity of the hydroxyl vibration located at about 1645 $cm^{-1}$ for 1F than 0F is indicative of the partial substation of fluoride for hydroxyl in the hydroxyapatite, forming fluorohydroxyapatite. The incorporation of fluoride enhances the chemical stability of apatite and inhibits its re-dissolution [31], justifying the higher biomineralization of 1F than the 0F sample. On the other hand, fluorite ($CaF_2$) is detected in the immersed 2F sample, via the absorption mode of Ca-F at about 420 $cm^{-1}$. The formation of fluorite consumes Ca, as shown below by the ICP analysis, and thereby limits the *in vitro* deposition of apatite, lowering the apatite-forming ability of 2F in comparison to 0F and 1F. In this regard, the preferred incorporation of fluoride into fluorite, rather than apatite, depletes the latter form fluoride, increasing the intensity of the hydroxyl vibration of 1645 $cm^{-1}$ compared to 1F. Note that the above FTIR assignments are compatible with Refs. [32-34].

The *in vitro* biodegradation of the samples was also studied by the ICP analysis on the SBF after soaking, in terms of the concentrations of Si, Mg, Ca and P ions, to deliberate the apatite-forming mechanism of the materials (Fig. 9). For all of the samples, the concentrations of Si and Mg ions are relatively enhanced with the soaking time due to the dissolution of the samples toward the SBF, where they do not directly take part in the structure of deposited apatite. Comparatively, the higher concentrations of these species for 0F over the entire range of incubation period can be because of the fact that this sample is F-free, whereas the degradation of 1F and 2F is retarded due to the high chemical affinities of F with both Si and Mg. However, the dissolution of 1F, in terms of Si and Mg ions, is a little lower than 2F (Figs. 9a and 9b) typically for the longer incubation periods, despite having the lower percentage of F. It can be attributed to the formation of the relatively-dense and continuous apatite layer on the incubated 1F surface (Figs. 6d and 6e), which retards the more dissolution of the underlying species. Regarding the Ca species, it is noticeable that its dissolution from the samples into the SBF, as evident in Fig. 9c from its higher concentrations in the soaked conditions compared to the fresh SBF, is critical for an effective deposition of apatite via creating a supersaturated solution. Although the Ca species is one of the principal components of apatite, the evaluation of the Ca content in the SBF gives no clear criterion to compare the biomineralization of the samples. Because a compromise between the dissolution of the samples and the deposition of apatite determines the content of Ca ions in the SBF after soaking the samples. However, the lower value of Ca in the SBF for 2F than 0F and 1F can be due to its limited dissolution (because of the high affinity with F) and the formation of fluorite on its surface consuming Ca, as pointed out above by the FTIR analysis. In contrast to the previous ions in the SBF, the amount of P ions is progressively reduced with immersion time for all of the samples (Fig. 9d). This is due to the progress of the apatite precipitation characterized above by SEM and FTIR, because the only reason for the changes in the P content is the deposition of apatite. Also, at any given immersion period, the concentration of P ions is ranked as 2F > 0F > 1F, suggesting that 1F and 2F have the maximum and minimum biomineralization abilities, respectively. In conclusion, the study of biodegradation by the ICP method on the SBF confirms the apatite-forming ability ranking of the samples found by the SEM, EDS and FTIR studies.

The pH value of the SBF after soaking the samples was also measured to establish another correlation between the bioactivity and biodegradation behaviors. The variations in pH can be explained by considering the ion-exchange reactions controlling the biomineralization, as pointed out by the ICP analysis. The fresh SBF has the pH value of 7.4. For all of the samples, pH is initially enhanced and then declined by increasing the immersion time, as represented in Fig. 10. The first enhancement is due to the dissolution of the cations from the samples into the SBF, which was above indicated in the ICP analysis (Fig. 9). Afterwards, the created vacancies surrounded by the lattice anions absorbs $H^+$ from the SBF, which thereby enhances pH. The following decrease in pH is indicative of the domination of the hydroxyapatite precipitation which demands hydroxyl of the SBF. As can be also seen, the pH value of the SBF for the F-containing samples (1F and 2F) is lower than that for 0F over the entire range of immersion period. This is because of the fact that fluoride is released from 1F and 2F, is replaced with hydroxyl anions of the SBF and thereby drops pH, which is in agreement with Ref. [26].

## 4. Concluding remarks

1. To obtain a homogenous diopside structure, the upper limit of the fluoride incorporation into the oxyfluorosilicate with Ca:Mg:Si = 1:1:2 was 2 mol%.

2. In the homogenous Ca-Mg oxyfluorosilicates, the lattice volume and crystallize size of diopside were enhanced with the fluoride addition, whereas they were reduced in the multiphase fluoridated samples.

3. The maximum *in vitro* apatite-forming ability among the single-phase silicates containing 0, 1 and 2 mol% F belonged to diopside doped with 1 mol% F.

4. The incorporation of F into hydroxyapatite deposited *in vitro* on the optimal sample was responsible for its better bioactivity, in comparison to the pure diopside sample.

5. The *in vitro* precipitation of fluorite on the surface of the 2 % F-doped sample declined its biomineralization than the optimal sample.

6. The biodegradation of the oxyfluorosilicates run some ion-exchange reactions which led to the formation of apatite on the surfaces.

## References

[1] L.L. Hench, Bioceramics: from concept to clinic, Journal of the American Ceramic Society, 74 (1991) 1487-1510.
[2] M. Vollenweider, T.J. Brunner, S. Knecht, R.N. Grass, M. Zehnder, T. Imfeld, W.J. Stark, Remineralization of human dentin using ultrafine bioactive glass particles, Acta Biomaterialia, 3 (2007) 936-943.
[3] Q. Du Min, Z. Bian, H. Jiang, D.C. Greenspan, A.K. Burwell, J. Zhong, B.J. Tai, Clinical evaluation of a dentifrice containing calcium sodium phosphosilicate (novamin) for the treatment of dentin hypersensitivity, American journal of dentistry, 21 (2008) 210-214.
[4] D. Gillam, J. Tang, N. Mordan, H. Newman, The effects of a novel Bioglass® dentifrice on dentine sensitivity: a scanning electron microscopy investigation, Journal of oral rehabilitation, 29 (2002) 305-313.
[5] V.A. Dubok, Bioceramics—Yesterday, Today, Tomorrow, Powder Metallurgy and Metal Ceramics, 39 (2000) 381-394.
[6] W. Xue, X. Liu, X. Zheng, C. Ding, Plasma-sprayed diopside coatings for biomedical applications, Surface and Coatings Technology, 185 (2004) 340-345.
[7] E. Salahinejad, R. Vahedifard, Deposition of nanodiopside coatings on metallic biomaterials to stimulate apatite-forming ability, Materials & Design, 123 (2017) 120-127.
[8] R. Vahedifard, E. Salahinejad, Microscopic and spectroscopic evidences for multiple ion-exchange reactions controlling biomineralization of CaO. MgO. 2SiO 2 nanoceramics, Ceramics International, (2017).
[9] Y. Miake, T. Yanagisawa, Y. Yajima, H. Noma, N. Yasui, T. Nonami, High-resolution and analytical electron microscopic studies of new crystals induced by a bioactive ceramic (diopside), Journal of dental research, 74 (1995) 1756-1763.
[10] J.W. Nicholson, Adhesive dental materials—a review, International journal of adhesion and adhesives, 18 (1998) 229-236.
[11] L. Seppä, Fluoride Release and Effect on Enamel Softening by Fluoride-Treated and Fluoride-Untreated Glass Ionomer Specimens, Caries research, 28 (1994) 406-408.
[12] J.A. Cury, L.M.A. Tenuta, Enamel remineralization: controlling the caries disease or treating early caries lesions?, Brazilian oral research, 23 (2009) 23-30.
[13] J. Cury, L. Tenuta, How to maintain a cariostatic fluoride concentration in the oral environment, Advances in dental research, 20 (2008) 13-16.
[14] M.J. Baghjeghaz, E. Salahinejad, Enhanced sinterability and in vitro bioactivity of diopside through fluoride doping, Ceramics International, 43 (2017) 4680-4686.
[15] E. Salahinejad, M. Hadianfard, D. Macdonald, I. Karimi, D. Vashaee, L. Tayebi, Aqueous sol–gel synthesis of zirconium titanate (ZrTiO 4) nanoparticles using chloride precursors, Ceramics International, 38 (2012) 6145-6149.
[16] E. Salahinejad, M. Hadianfard, D. Macdonald, M. Mozafari, D. Vashaee, L. Tayebi, Zirconium titanate thin film prepared by an aqueous particulate sol–gel spin coating process using carboxymethyl cellulose as dispersant, Materials Letters, 88 (2012) 5-8.
[17] S. Rastegari, O.S.M. Kani, E. Salahinejad, S. Fadavi, N. Eftekhari, A. Nozariasbmarz, L. Tayebi, D. Vashaee, Non-hydrolytic sol-gel processing of chloride precursors loaded at forsterite stoichiometry, Journal of Alloys and Compounds, 688 (2016) 235-241.

[18] E. Salahinejad, M. Hadianfard, D. Macdonald, M. Mozafari, K. Walker, A.T. Rad, S. Madihally, D. Vashaee, L. Tayebi, Surface modification of stainless steel orthopedic implants by sol–gel ZrTiO4 and ZrTiO4–PMMA coatings, Journal of biomedical nanotechnology, 9 (2013) 1327-1335.
[19] E. Salahinejad, M. Hadianfard, D. Macdonald, M. Mozafari, D. Vashaee, L. Tayebi, Multilayer zirconium titanate thin films prepared by a sol–gel deposition method, Ceramics International, 39 (2013) 1271-1276.
[20] M. Mozafari, E. Salahinejad, V. Shabafrooz, M. Yazdimamaghani, D. Vashaee, L. Tayebi, Multilayer bioactive glass/zirconium titanate thin films in bone tissue engineering and regenerative dentistry, Int J Nanomedicine, 8 (2013) 1665-1672.
[21] P. Rouhani, E. Salahinejad, R. Kaul, D. Vashaee, L. Tayebi, Nanostructured zirconium titanate fibers prepared by particulate sol–gel and cellulose templating techniques, Journal of Alloys and Compounds, 568 (2013) 102-105.
[22] E. Salahinejad, M. Hadianfard, D. Vashaee, L. Tayebi, Effect of precursor solution pH on the structural and crystallization characteristics of sol–gel derived nanoparticles, Journal of Alloys and Compounds, 589 (2014) 182-184.
[23] M. Mozafari, E. Salahinejad, S. Sharifi-Asl, D. Macdonald, D. Vashaee, L. Tayebi, Innovative surface modification of orthopaedic implants with positive effects on wettability and in vitro anti-corrosion performance, Surface Engineering, 30 (2014) 688-692.
[24] T. Kokubo, H. Takadama, How useful is SBF in predicting in vivo bone bioactivity?, Biomaterials, 27 (2006) 2907-2915.
[25] D.S. Brauer, N. Karpukhina, R.V. Law, R.G. Hill, Structure of fluoride-containing bioactive glasses, Journal of Materials Chemistry, 19 (2009) 5629-5636.
[26] D.S. Brauer, N. Karpukhina, M.D. O'Donnell, R.V. Law, R.G. Hill, Fluoride-containing bioactive glasses: effect of glass design and structure on degradation, pH and apatite formation in simulated body fluid, Acta Biomaterialia, 6 (2010) 3275-3282.
[27] M. ClrunnoN, S. SuBNo, C. Pnnwrrr, J. Plpmn, High-Temperature Grystal Ghemistry of Acmite, Diopside, Hedenbergite, Jadeite, Spodumene, and Ureyite, American Mineralogist, 58 (1973) 594-618.
[28] A. Wilson, J. Nicholson, Acid–Base Cements: Their Biological and Other Applications, Cambridge University Press: Cambridge, 1993.
[29] A. Rafferty, A. Clifford, R. Hill, D. Wood, B. Samuneva, M. Dimitrova-Lukacs, Influence of Fluorine Content in Apatite–Mullite Glass-Ceramics, Journal of the American Ceramic Society, 83 (2000) 2833-2838.
[30] D.S. Brauer, R.G. Hill, M.D. O'Donnell, Crystallisation of fluoride-containing bioactive glasses, Physics and Chemistry of Glasses-European Journal of Glass Science and Technology Part B, 53 (2012) 27-30.
[31] P. Taddei, E. Modena, A. Tinti, F. Siboni, C. Prati, M.G. Gandolfi, Effect of the fluoride content on the bioactivity of calcium silicate-based endodontic cements, Ceramics International, 40 (2014) 4095-4107.
[32] I. Rehman, W. Bonfield, Characterization of hydroxyapatite and carbonated apatite by photo acoustic FTIR spectroscopy, Journal of Materials Science: Materials in Medicine, 8 (1997) 1-4.
[33] A. Ślósarczyk, Z. Paszkiewicz, C. Paluszkiewicz, FTIR and XRD evaluation of carbonated hydroxyapatite powders synthesized by wet methods, Journal of Molecular Structure, 744 (2005) 657-661.
[34] K. Tahvildari, S. Ghammamy, H. Nabipour, CaF2 nanoparticles: synthesis and characterization, International Journal of Nano Dimension, 2 (2012) 269-273.

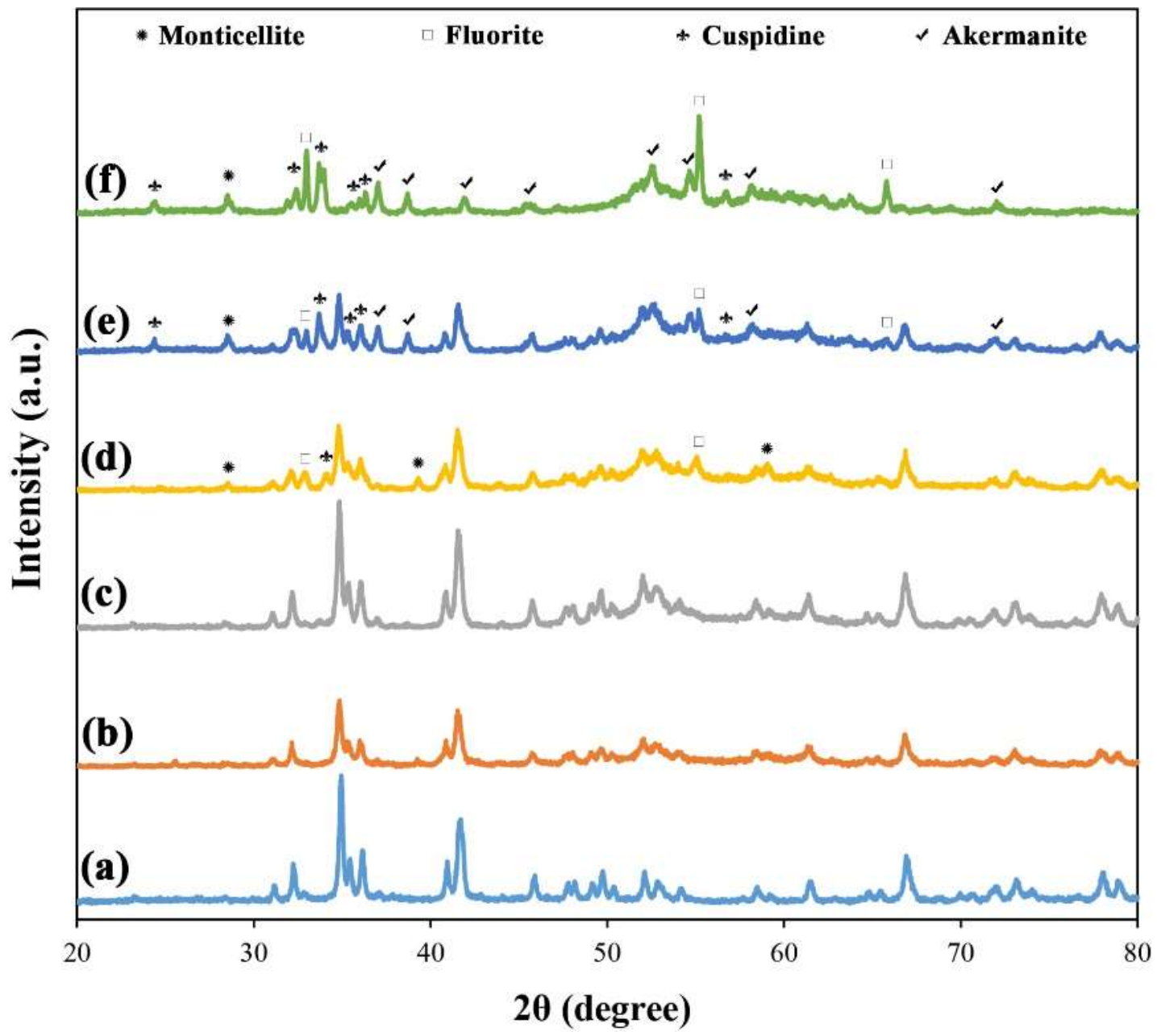

Fig. 1. XRD patterns of the 0F (a), 1F (b), 2F (c), 3F (d), 4F (e) and 5F (f) samples before soaking in the SBF.

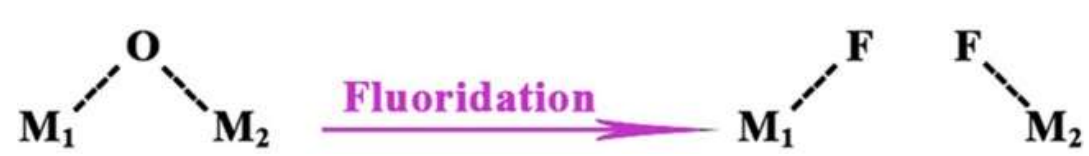

Fig. 2. Transformation of bridging-to-nonbridging (oxide-to-fluoridated) bonds due to fluoridation ($M_1$ and $M_2$ can be Si, Ca and Mg).

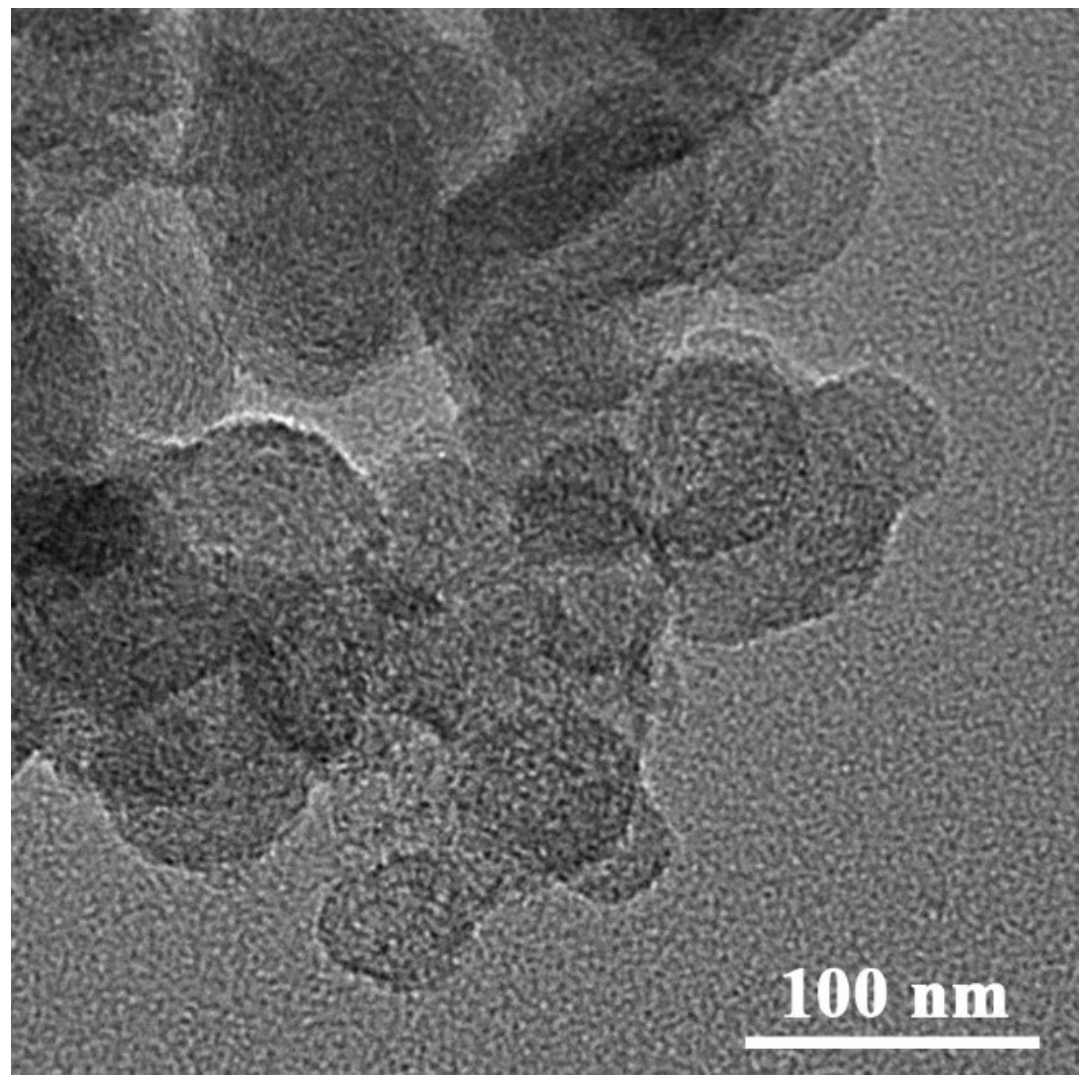

Fig. 3. TEM micrographs of the 0F powder sample.

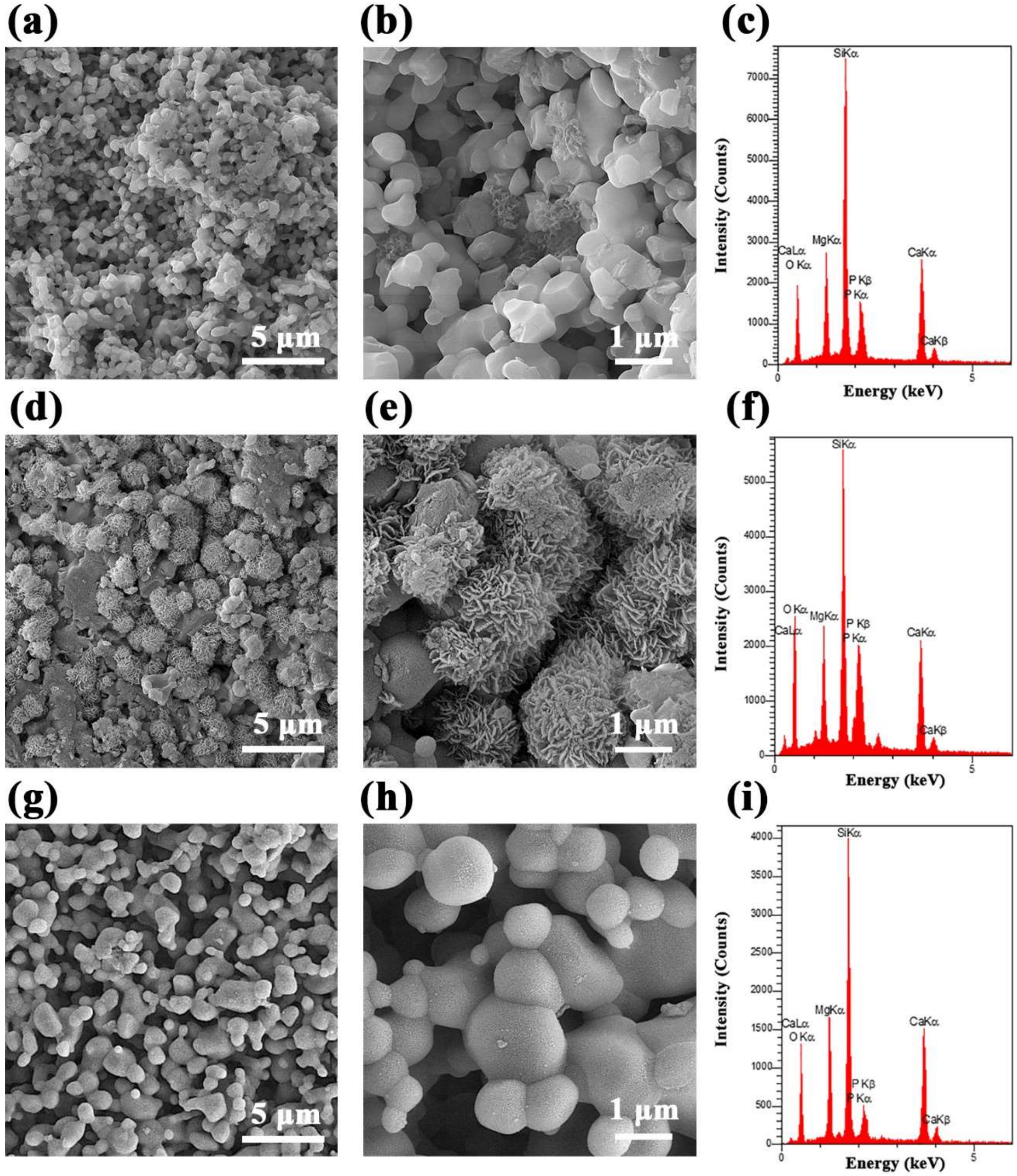

Fig. 4. SEM micrographs (in two magnifications) and EDS spectra of the 0F (a, b, c), 1F(d, e, f) and 2F (g, h, i) samples after 3 days of soaking in the SBF.

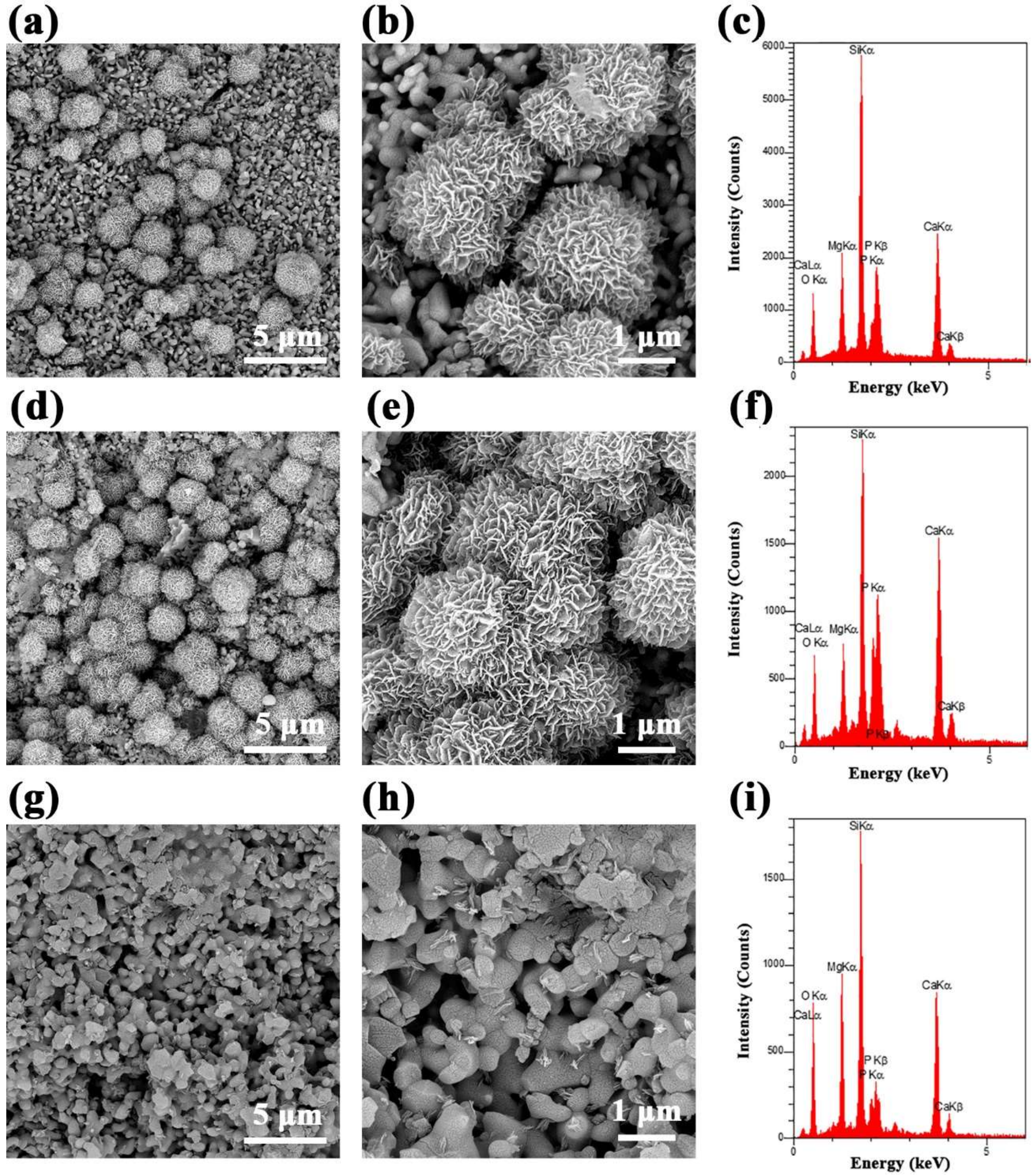

Fig. 5. SEM micrographs (in two magnifications) and EDS spectra of the 0F (a, b, c), 1F(d, e, f) and 2F (g, h, i) samples after 7 days of soaking in the SBF.

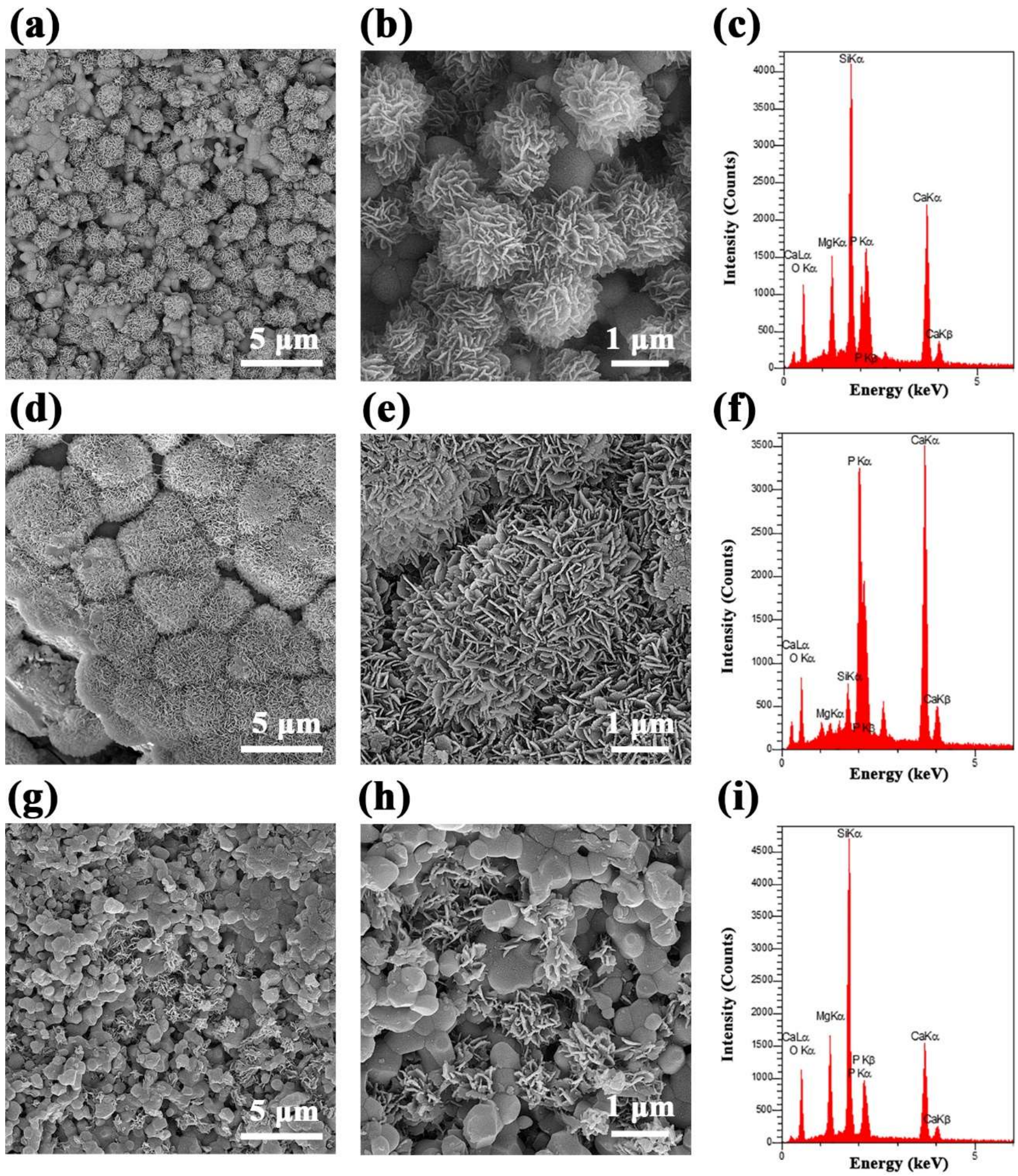

Fig. 6. SEM micrographs (in two magnifications) and EDS spectra of the 0F (a, b, c), 1F(d, e, f) and 2F (g, h, i) samples after 14 days of soaking in the SBF.

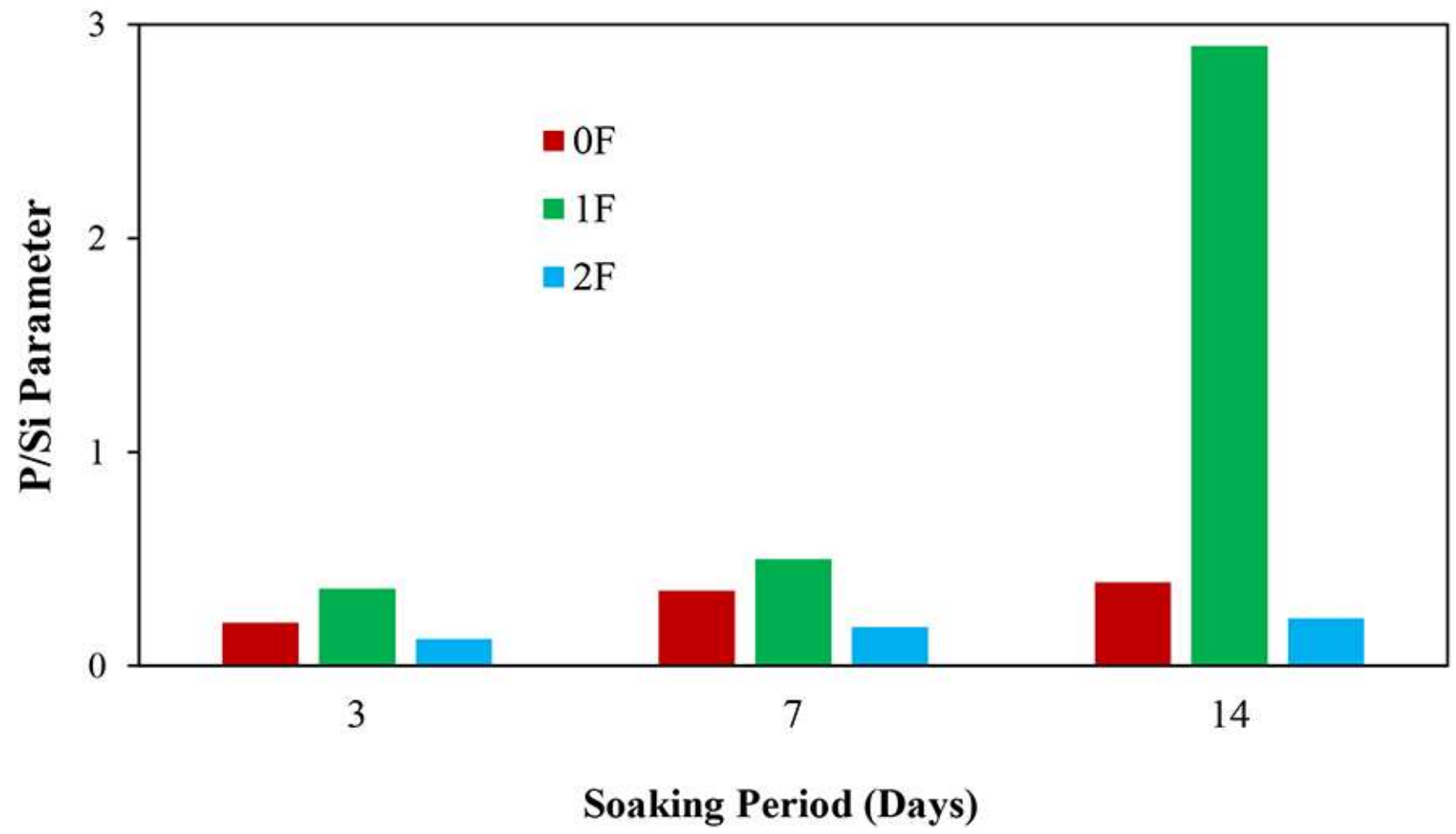

Fig. 7. P/Si parameters for the samples incubated in the SBF for the different periods, extracted from the EDS spectra.

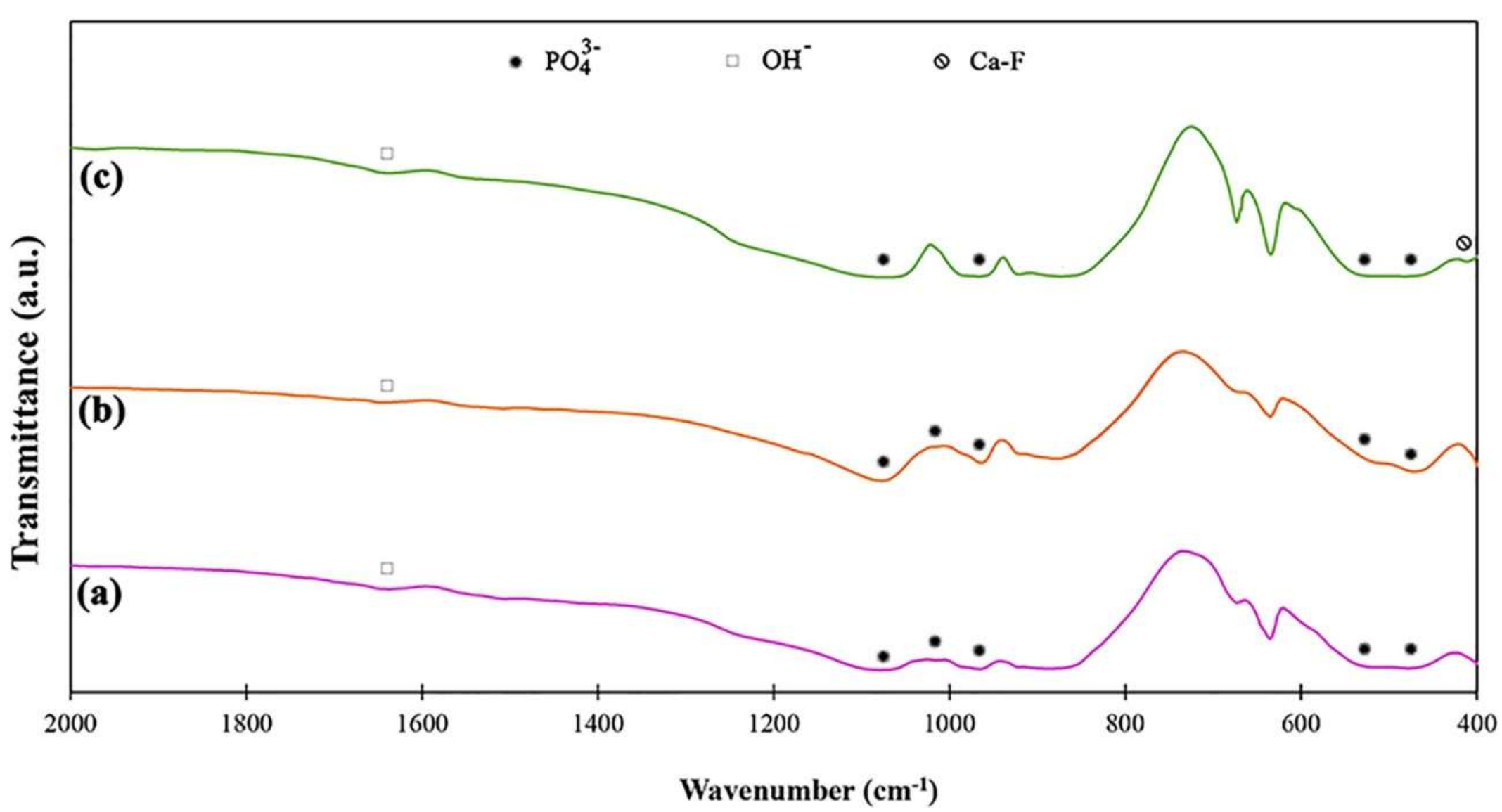

Fig. 8. FTIR spectra of the 0F (a), 1F(b) and 2F (c) samples after soaking in the SBF.

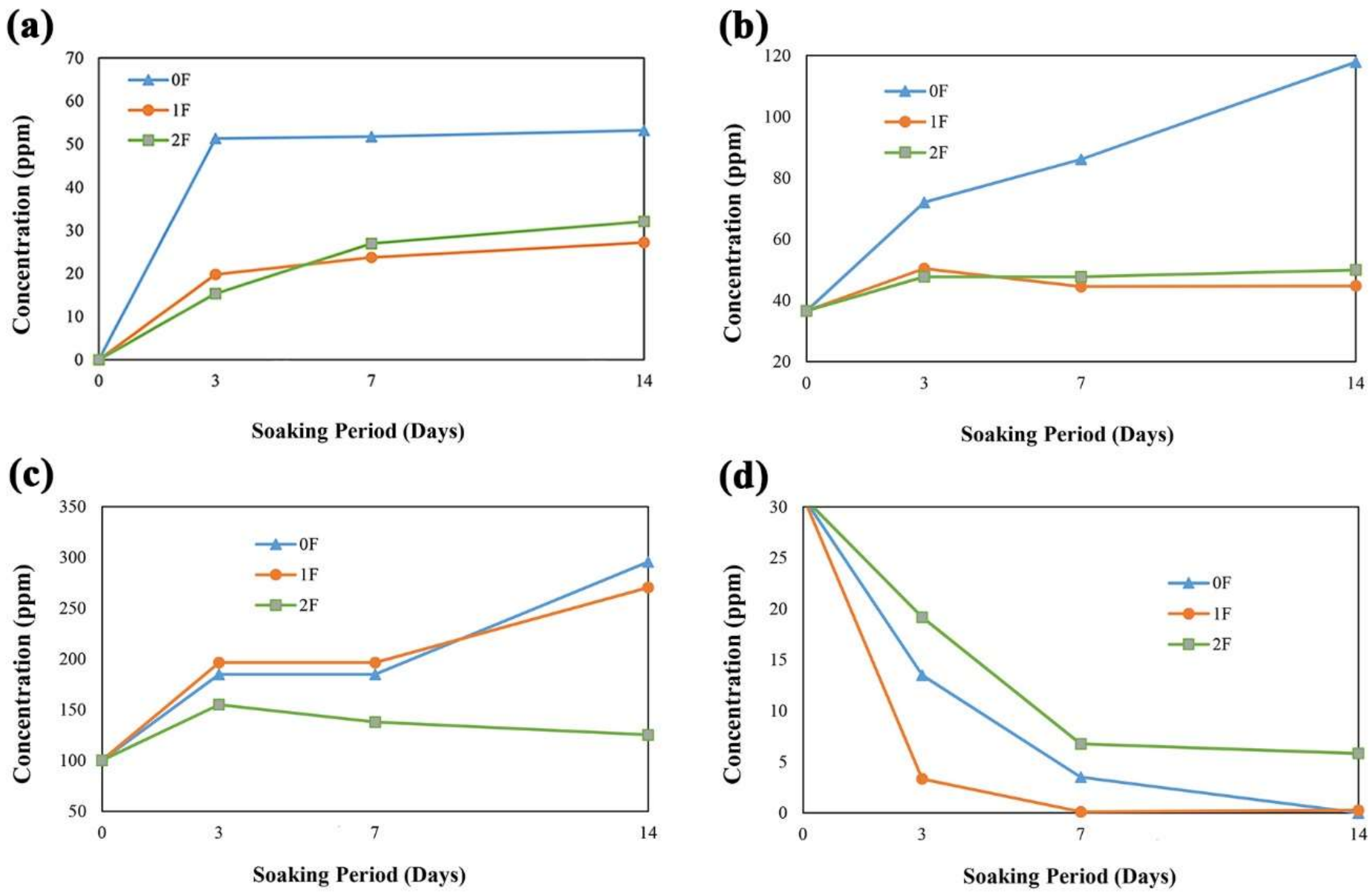

Fig. 9. ICP concentrations of Si (a), Mg (b), Ca (c) and P (d) ions in the SBF after immersion of the 0F, 1F and 2F samples for the different days.

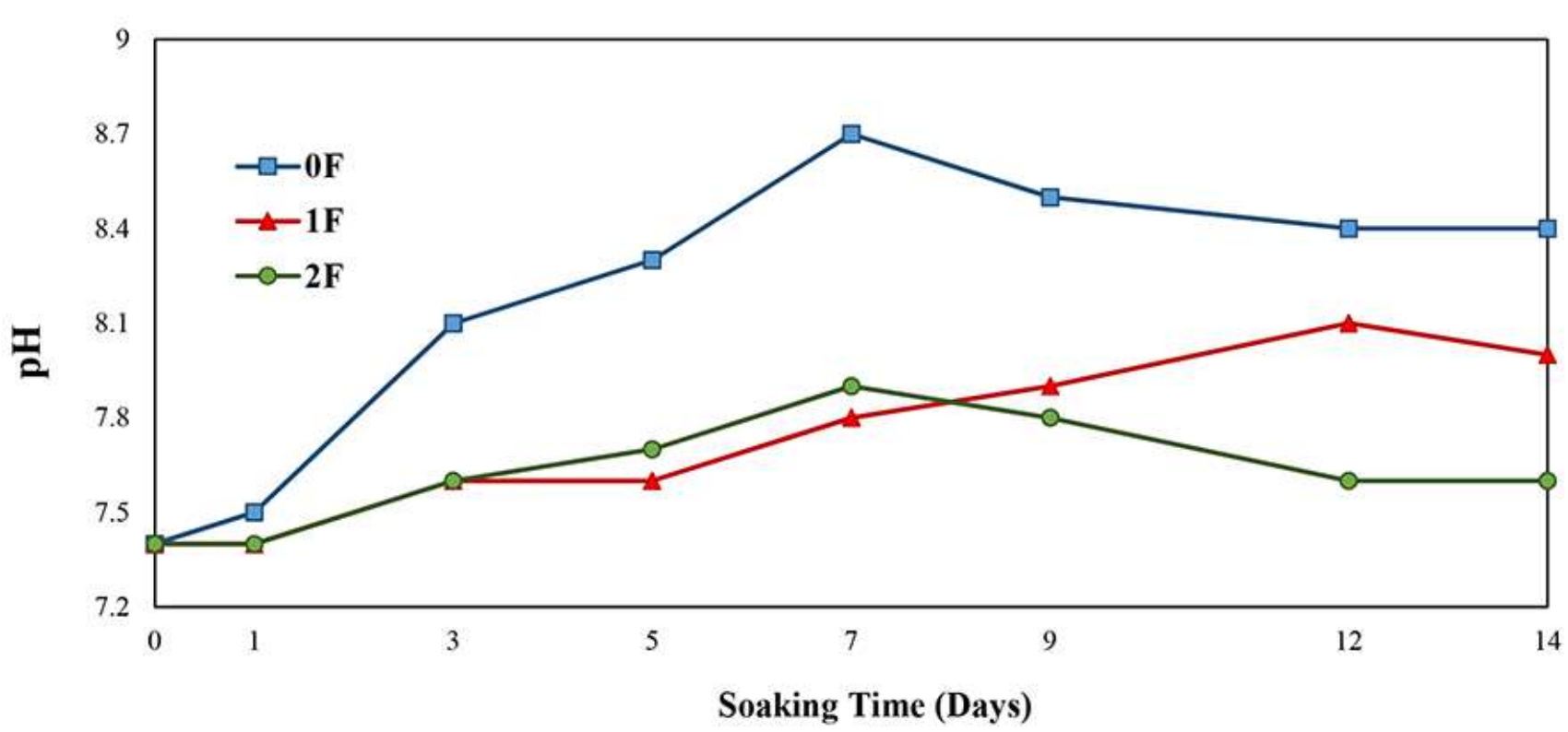

Fig. 10. pH variations of the SBF after immersion of the samples for the various durations.

## Table

Table 1. Lattice parameters, interaxial angle, lattice volume and crystallite size of diopside doped with fluoride at the different levels (the last digit of the calculated numbers has been removed due to uncertainty).

| Sample | *a* (Å) | *b* (Å) | *c* (Å) | *β* (°) | *V* (Å$^3$) | *D* (nm) |
|---|---|---|---|---|---|---|
| **Theory [27]** | 9.746 | 8.899 | 5.251 | 105.6 | 438.641 | - |
| **0F** | 9.687 | 8.841 | 5.250 | 106.3 | 431.552 | 57 |
| **1F** | 9.758 | 8.933 | 5.249 | 105.9 | 440.041 | 50 |
| **2F** | 9.752 | 8.946 | 5.248 | 105.7 | 440.762 | 35 |
| **3F** | 9.763 | 8.857 | 5.248 | 105.9 | 436.438 | 42 |
| **4F** | 9.700 | 8.868 | 5.25 | 105.9 | 434.325 | 53 |